\documentclass[twocolumn,showpacs,preprintnumbers,amsmath,amssymb,pra]{revtex4}

\usepackage{graphicx}
\usepackage{dcolumn}
\usepackage{bm}

\def\>{\rangle}
\def\<{\langle}

\makeatletter
\newcommand{\Rmnum}[1]{\expandafter\@slowromancap\romannumeral #1@}
\makeatother

\begin{document}

\newcommand{\ket}[1]{| #1 \rangle}
\newcommand{\bra}[1]{\langle #1 |}

\preprint{}

\title{Entanglement-Seeded-Dual Optical Parametric Amplification:\\
Applications to Quantum Communication, Imaging, and Metrology}

\author{Ryan T. Glasser}
 \email{rglass1@lsu.edu}
 \affiliation{%
Horace C. Hearne Jr. Institute for Theoretical Physics,\\
Department of Physics and Astronomy, Louisiana State University, Baton Rouge LA 70803.
}
\author{Hugo Cable}%
 \email{hcable@phys.lsu.edu}
 \affiliation{%
Horace C. Hearne Jr. Institute for Theoretical Physics,\\
Department of Physics and Astronomy, Louisiana State University, Baton Rouge LA 70803.
}%
 \author{F. De Martini}
 \affiliation{%
 University of Rome ``La Sapienza"\\
 Piazzale Aldo Moro, 2 00185 Roma-Italy
 }%
 \author{Fabio Sciarrino}
 \affiliation{%
 University of Rome ``La Sapienza"\\
 Piazzale Aldo Moro, 2 00185 Roma-Italy
 }%
 \author{Chiara Vitelli}
 \affiliation{%
 University of Rome ``La Sapienza"\\
 Piazzale Aldo Moro, 2 00185 Roma-Italy
 }%
 \author{Jonathan P. Dowling}%
\affiliation{%
Horace C. Hearne Jr. Institute for Theoretical Physics,\\
Department of Physics and Astronomy, Louisiana State University, Baton Rouge LA 70803.
}%

\date{\today}

\begin{abstract}
The study of optical parametric amplifiers (OPAs) has been successful in describing and
creating nonclassical light for use in fields such as quantum metrology and quantum
lithography [Agarwal, {\it et al.}, J. Opt. Soc. Am. B, {\bf 24}, 2 (2007)].  In this
paper we present the theory of an OPA scheme utilizing an entangled state input.  The
scheme involves two identical OPAs seeded with the maximally path-entangled
$\ket{\rm{N00N}}$ state $(\ket{2,0}+\ket{0,2})/\sqrt{2}$.  The stimulated amplification
results in output state probability amplitudes that have a dependence on the number of
photons in each mode, which differs greatly from two-mode squeezed vacuum.  The output
contains a family of entangled states directly applicable to quantum key distribution.
Specific output states allow for the heralded creation of $N=4$ $N00N$ states, which may
be used for quantum lithography, to write sub-Rayleigh fringe patterns, and for quantum
interferometry, to achieve Heisenberg-limited phase measurement sensitivity.
\end{abstract}


\maketitle

\section{Introduction}

Nonclassical states of light have been studied in depth both experimentally and
theoretically since the emergence of quantum electronics.  Squeezed light, in
particular, has been applied to a variety of systems, including interferometry,
lithography, and cryptography which show improvement beyond limitations imposed by
classical optics \cite{agarwal1,caves81,grangier4,boto}.  One such device that creates a
type of squeezed light is an optical parametric amplifier (OPA).   OPAs are typically
non-centrosymmetric crystals that exhibit a nonzero $\chi^{(2)}$ optical nonlinearity
\cite{"boydbook"}.
Pump, signal, and idler modes propagate through the crystal, and photons from the pump
beam are down converted into lower energy photons in the signal and idler modes.
Previous work focused on the case that the signal and idler modes couple to the vacuum
at the input. This produces the two-mode squeezed vacuum state, which exhibits a highly
nonclassical behavior \cite{yuen2,caves81,grangier4}.  In the present paper we analyze a
scheme in which two identical OPAs are seeded by entangled photon pairs.  The scheme
produces a heralded source for a large family of entangled states, of interest for
applications in quantum information, metrology, and imaging.  These states are generated
by conditioning the output on photodetection on two of the four total output modes.

A particularly useful heralded state that our scheme generates is the so-called `N00N'
state with $N=4$.  A N00N state is a maximally path entangled state such that, in a
Fock-state basis,
$\ket{\rm{N00N}}\propto\ket{N}_{A}\ket{0}_{B}+e^{iN\varphi}\ket{0}_{A}\ket{N}_{B}$,
where $\varphi$ is the relative phase difference between the two spatial modes A and B
\cite{boto}.  These states allow for super-resolution by producing lithographic features
with a minimum size of $\lambda/(2N)$, when incident on an $N$-photon absorbing
substrate, thus allowing an $N$-fold enhancement over standard lithographic methods
\cite{boto,agarwal1,sciarrino}.  N00N states have also been shown to exhibit
super-sensitivity in interferometric applications, thus reaching the Heisenberg Limit of
$\Delta\phi=1/N$ \cite{agarwal1,holland,durkin,kapale}.  Classically, in an
interferometer using coherent light, precision in phase-uncertainty measurement is
limited by the shot-noise limit of $\Delta\phi=1/\sqrt{\bar{n}}$, where $\bar{n}$ is the
average photon number.  Experimentally, up to $N=4$ $N00N$ states have been reported and
shown to exhibit both super-sensitivity and super-resolution
\cite{zeilinger,obrienscience}. However, implementing N00N-state generators that produce
states of photon number greater than two, which simultaneously achieve high fidelities
and high flux, is very challenging experimentally.  Recently we proposed a scheme that
scales well with $N$ and works for an input of any superposition of $\ket{N,N}$ photons
coupled with feed-forward \cite{hugo}.  Our new scheme, presented here, produces
heralded $N=4$ $N00N$ states with relatively high probability, and is experimentally
accessible.  Additionally, a wide variety of other useful entangled states are produced
in our new scheme, which can be applied to quantum metrology and cryptography.

In section \Rmnum{2} we will review the process of optical parametric amplification and
squeezing.  In section \Rmnum{3} we describe the novel entanglement-seeded-dual optical
parametric amplification scheme.  Finally, in section \Rmnum{4} we analyze the
properties of the output state, including probabilities and applications.

\section{Optical Parametric Amplification}

To obtain the input state for our scheme, some squeezing formalism will be reviewed.  We
will work in the Heisenberg picture and use a Fock (number) state basis throughout the
paper.  Modes are represented with capital letters, such as mode A, mode B, and so on.
The creation and annihilation operators for the respective modes are $\hat{a}^{\dag}$,
$\hat{a}$, $\hat{b}^{\dag}$, and $\hat{b}$. The mode labels are dropped from the kets,
but proceed in alphabetical order such that $\ket{N}_{A}\ket{M}_{B}\equiv\ket{N,M}$.

The unitary operator describing the action of an OPA is the two-mode squeezing operator
\cite{drummondbook},
\begin{eqnarray}
\hat{S}(\xi)=e^{-\xi\hat{a}^{\dag}\hat{b}^{\dag}+\xi^{*}\hat{a}\hat{b}},
\end{eqnarray}
where $\xi=re^{i\varphi}$ is the complex squeezing parameter.  Here a strong, undepleted
classical pump is also assumed.  Here $r$ is the gain and $\varphi$ is the phase
associated with the OPA.  As previously mentioned, the gain $r$ depends on the pump amplitude and the length
and nonlinearity of the crystal.  The action of the two-mode squeezing operator on a
vacuum input produces the two-mode squeezed vacuum \cite{yuen1},
\begin{eqnarray}
\hat{S}(\xi)\ket{0,0}=\frac{1}{\cosh{r}}\sum_{n=0}^{\infty}(-1)^{n}
e^{in\varphi}\tanh^{n}{(r)}\ket{n,n}.
\end{eqnarray}

In the low-gain limit, an OPA acts as a spontaneous parametric down convertor (SPDC)
\cite{"boydbook"}.  The output is then well approximated as vacuum and a stream of
$\ket{1,1}$ states.  We assume spontaneous parametric downconversion to initially
produce the state $\ket{1,1}$, then input this two-mode state on a 50:50 beam splitter
that takes modes $\hat{a}^{\dag}\rightarrow(\hat{a}^{\dag}+i\hat{b}^{\dag})/\sqrt{2}$
and $\hat{b}^{\dag}\rightarrow(i\hat{a}^{\dag}+\hat{b}^{\dag})/\sqrt{2}$. Due to the
Hong-Ou-Mandel effect, two, single, indistinguishable photons that are incident
simultaneously on a beam splitter evolve to a superposition in which only one mode is
occupied by both photons at the output. We then obtain our desired entangled low-N00N
input state $(\ket{2,0}+\ket{0,2})/\sqrt{2}$ \cite{mandel}.

\begin{figure}[t]
\includegraphics[width=8.5cm]{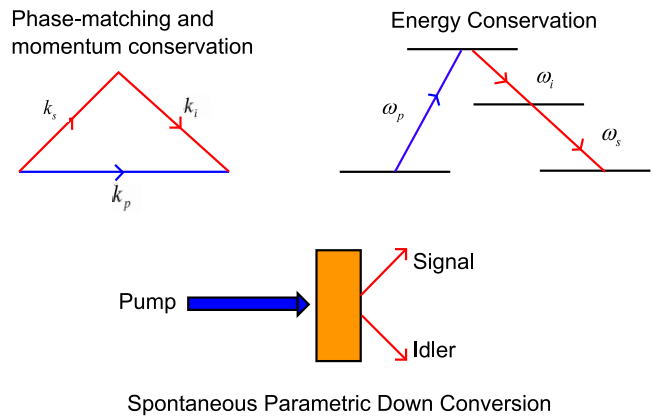}
\caption{\label{fig:epsart1} Schematic of optical parametric amplification and
spontaneous parametric downconversion.  The bottom diagram shows the crystal being
pumped by a laser, along with the signal and idler output modes.  The top left diagram
depicts momentum conservation between the pump, signal and idler modes, otherwise known
as the phase-matching condition. The top right picture diagramatically shows energy
conservation of the system, such that the frequency of the pump is equal to the signal
and idler frequencies added together.  These restrictions lead to the path entanglement
we desire in our scheme.}
\end{figure}

The action of the unitary operator describing an OPA transforms input modes $A$ and $B$
as \cite{caves81},
\begin{subequations}
\label{eq:whole}
\begin{eqnarray}
\hat{S}(\xi)\hat{a}^{\dag}\hat{S}^{\dag}(\xi)=\hat{a}^{\dag}\cosh{r}+\hat{b}e^{-i\varphi}\sinh{r},
\label{subeq:1}
\end{eqnarray}
\begin{eqnarray}
\hat{S}(\xi)\hat{b}^{\dag}\hat{S}^{\dag}(\xi)=\hat{b}^{\dag}\cosh{r}+\hat{a}e^{-i\varphi}\sinh{r}.
\label{subeq:2}
\end{eqnarray}
\end{subequations}
In any type of parametric amplification, energy and momentum must be conserved.
Momentum conservation provides a phase-matching condition between the pump and the
signal and idler modes, which generates path entanglement.  Conservation of energy
requires the frequencies of the signal and idler to add up to the frequency of the pump
beam.  We will be considering the case of degenerate parametric amplification, such that
the signal $\omega_{s}$ and idler $\omega_{i}$ frequencies are identical and half of the
pump frequency $\omega_{p}$; that is
$\frac{1}{2}\omega_{p}=\omega_{s}=\omega_{i}=\omega$ \cite{"boydbook"}.  Schematically,
energy and momentum conservation can be understood from Fig.~\ref{fig:epsart1}.

Much of the research involving OPAs and two-mode squeezing assumed vacuum input modes
\cite{agarwal1,demartini,yurke1,onohofmann1}.  This results in the two-mode-squeezed
vacuum state previously mentioned.  Some theoretical and experimental work has assumed
non-vacuum inputs, typically with coherent light input in one mode
\cite{kumar1,heidmann1,grangier2,grangier3,grangier4}. Indeed, arguably the most useful
limit is the low-gain limit of an OPA, which produces (to a good approximation)
spontaneous parametric downconversion, that is, the $\ket{1,1}$ state.  This state, and
more generally two-mode squeezed vacuum, have been used to help beat the shot-noise
limit in interferometric applications \cite{italian}. De Martini's group has recently
demonstrated the idea of seeding OPAs with nonclassical light, namely number states.
They showed that entanglement was preserved between two of the output modes and one
input mode of an OPA.  Importantly, the second input mode is detected as a trigger for
the experiment \cite{demartini}.  They described this process as quantum injection of an
OPA, where one of the input modes was seeded with one photon from a down-converted pair
produced by SPDC.

\section{Entanglement Seeded Optical Parametric Amplification}

\begin{figure}[t]
\includegraphics[width=8.5cm]{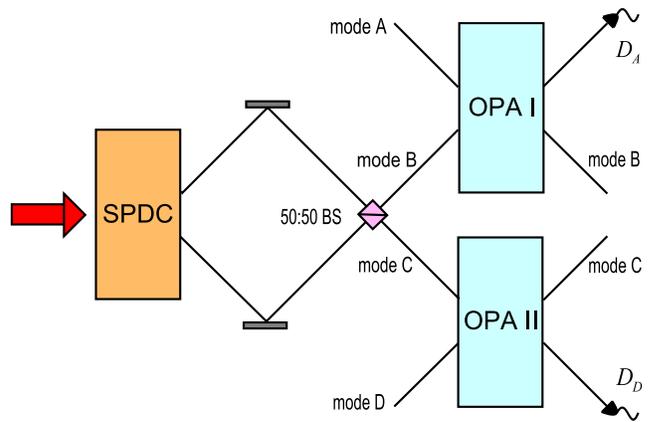}
\caption{\label{fig:epsart2} Diagram of our entanglement-seeded-dual optical
amplification scheme.  On the far left is the relatively weak pump beam pumping a
nonlinear crystal in the low-gain regime in order to produce spontaneous parametric
downconversion.  This output state, taken initially to be $\ket{1,1}$, is then incident
on a beamsplitter, which leads to the maximally spatially entangled state
$(\ket{2,0}+\ket{0,2})/\sqrt{2}$.  This state is then incident into one mode of each of
the two OPAs.  The other two modes are left as vacuum inputs.  We then assume that OPA I
and OPA II are pumped by the same high gain laser in order to achieve parametric
amplification.  Note that the pumps for all three of the nonlinear crystals are to be
phase locked.  All four modes are amplified, resulting in entanglement between modes B
and C.  We have drawn photodetectors at modes A and D which are to be used in heralded
production of specific states.}
\end{figure}

While optical parametric amplification of vacuum input states produces interesting
squeezed vacuum states, we consider a scheme in which highly nonclassical states of
light are amplified.  Rather than seeding an OPA setup with either vacuum modes or
number states, we assume an entangled number state input.  Our scheme involves two OPAs,
for a total of four input modes, which are seeded in two of the modes with the state
$(\ket{2,0}+\ket{0,2})/\sqrt{2}$.  These two modes are then fed into the dual OPA scheme
as modes B and  C, leaving vacuum input in modes A and D, as seen in
Fig.~\ref{fig:epsart2}.  With this notation it is transparent that the inner two modes
contain the entangled-state input.  Thus, the total input state may be written as
\begin{eqnarray}
\ket{\rm{input}}\propto\ket{0,2,0,0}+\ket{0,0,2,0},
\end{eqnarray}
where we drop the consecutive mode labels A, B, C, and D.  By assuming an entangled
input we are naturally led to various questions about the output state. First and
foremost, is the output state entangled?  Due to amplification, has the degree of
entanglement from the input state deteriorated, or has the path entanglement been
retained?  Also, what are the applications of the output state and with what
probabilities does a given state occur?

Armed with the total input state and the squeezing operator transformations, we
calculate the output of the scheme.  We carry out the calculation by rewriting the input
state in terms of the creation operators corresponding to the appropriate modes, which
initially contain photons.

The state is then subject to the two OPA transformations $\hat{S}_{1}(\xi)$ and
$\hat{S}_{2}(\xi)$.  It should be clear that both OPAs are assumed to have the same
complex squeezing parameter $\xi$, which experimentally means they have the same
$\chi^{(2)}$ nonlinearity, are the same length, and cut to have the same phase matching
condition (more simply, they are identical).  Due to the unitarity of the two-mode
squeezing operator we are able to resolve the identity and apply the operators to the
input state, thus resulting in the output state:
\begin{eqnarray}
\ket{\rm{output}}=\frac{1}{2}(\hat{S}_{1}\hat{b}^{\dag}\hat{S}_{1}^{\dag}\hat{S}_{1}\hat{b}^{\dag}\hat{S}_{1}^{\dag}\hat{S}_{1}\hat{S}_{2}\ket{0,0,0,0}+\nonumber\\
\hat{S}_{2}\hat{c}^{\dag}\hat{S}_{2}^{\dag}\hat{S}_{2}\hat{c}^{\dag}\hat{S}_{2}^{\dag}\hat{S}_{2}\hat{S}_{1}\ket{0,0,0,0}).
\end{eqnarray}

Each of the two-mode squeezing operators transforms only two of the input modes.
$\hat{S}_{1}$ acts only on modes A and B while $\hat{S}_{2}$ acts on modes C and D.
Additionally, the two unitary operators commute with one another due to the fact that
the different mode operators commute.  The total output state is then:
\begin{eqnarray}
&&\ket{\rm{output}}  =
C(r)\sum_{n=0}^{\infty}\sum_{m=0}^{\infty}(-e^{i\varphi}\tanh{r})^{n+m}\nonumber\\
&&\times(\kappa(n)\ket{n,n+2,m,m}+\kappa(m)\ket{n,n,m+2,m}).
\end{eqnarray}
Here, $\varphi$ is the phase associated with the two OPAs, $n$ is the index resulting
from the two-mode squeezing due to $\hat{S}_{1}$ and $m$ the index corresponding to the
squeezing induced by $\hat{S}_{2}$.  The constant $C(r)$ depends only on the gain of the
OPAs as $C(r)=\frac{1}{2}(\cosh{r})^{-4}$.  Also, $\kappa(n)=\sqrt{n+1}\sqrt{n+2}$.

We see immediately that the inner two modes, B and C, are path entangled just as the
input state was.  Modes A and D will prove to be particularly valuable when detected,
thus giving us information about the inner two modes.  The output of the scheme is
similar to a two two-mode squeezed vacuum state, with the entangled input state being
amplified in the inner two modes.

\section{Discussion}

The output state is particularly useful when we consider placing photodetectors $D_{A}$
and $D_{D}$ at the outputs of the transformed modes A and D.  If we assume perfect
number-resolving photodetectors which implement projective measurements on modes A and
D, we are able to determine with certainty which state the inner two entangled modes are
in.  This gives us a specific heralded entangled state depending on what photon numbers
we measure at $D_{A}$ and $D_{D}$.  The entangled state after detecting $n$ photons at
detector $D_{A}$ and $m$ photons at detector $D_{D}$ will then be
\begin{eqnarray}
\ket{\rm{heralded}}\propto\ket{n+2,m}+\ket{n,m+2}
\end{eqnarray}
in modes B and C.  The probability of detecting these $n$ and $m$ photons at their
respective detectors is given by
\begin{eqnarray}
\rm{Prob}(n,m) & = & C(r)^{2}\nonumber\tanh^{2(n+m)}{r}\nonumber\\
&&  \times[(n+1)(n+2)+(m+1)(m+2)] 
\end{eqnarray}

\begin{figure}[t]
\includegraphics[width=9cm]{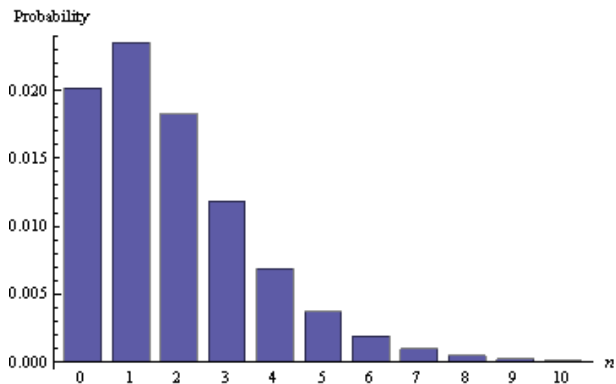}
\caption{\label{fig:epsart3} Probability of obtaining output states with $n=m$ for a
fixed gain of $r=1.08$. A joint detection of equal photon number at modes $a$ and $d$
results in the inner two modes being in the state $\ket{n+2,n}+\ket{n,n+2}$.  We see
that for this experimentally feasible gain that the vacuum term $n=0$ is no longer the
most probable outcome, whereas the desirable $n=1$ output is.}
\end{figure}

We can see that the parametric amplification results in an output state that is
dependent on the number of photons in the four modes.  For low values of gain, in which
we expect spontaneous parametric downconversion, the vacuum $n=m=0$ term dominates, due
to the exponential dependence on $n$ and $m$ of the hyperbolic tangent.  However, for
higher values of gain, in which we obtain parametric amplification, the amplified vacuum
term is no longer the most probable outcome, as seen in Fig.~\ref{fig:epsart3}.  The
maximum shifts towards states with higher photon numbers.  Additionally, the photon
number difference between the two inner modes is a defining characteristic, which makes
our heralded scheme nontrivial.

The immediate consequence of the photon number difference in the two inner modes of the
output state applies to quantum cryptography.  We imagine detecting $n$ photons at
$D_{A}$ and $m$ photons at $D_{D}$.  If we have perfect number resolving detectors, any
time we measure $n=m$ we have the inner mode entangled state
$(\ket{n+2,n}+\ket{n,n+2})/\sqrt{2}$.  To begin the QKD protocol, photodetector
measurements at $D_{A}$ and $D_{D}$ are announced publicly, while photon number
measurements afterwards on modes B and C by two parties (Alice and Bob) will be
perfectly correlated.  The time-energy entanglement of the two modes results in a
violation of the classical separability bound of the joint time and energy uncertainties
$(\Delta{E_{B,C})}^{2}(\Delta{t_{B,C})}^{2}\geq\hbar^{2}$
\cite{howellqkd,khanhowell}.  This type of entanglement is exploited to create a
one-time pad.  A setup analogous to the experiment carried out by Howell's group can
then be implemented \cite{howellqkd}.  In their scheme arrival times of photon pairs
created from SPDC, which are highly correlated, are used to create a cryptographic key.
A time-bin setup is used in order to ensure that detections at both Alice and Bob's
positions are due to the same SPDC pair.  This discretization of continuous-variable
entanglement has been implemented experimentally \cite{howellqkd}.

In our scheme, we exploit the number difference between the two modes, as well as the
time-energy entanglement, in order to create a key.  After the values measured at
$D_{A}$ and $D_{D}$ are publicly announced, one of each of the remaining modes is sent
to Alice, and the other to Bob.  Each of them then makes a photon number measurement on
the mode they have received.  The analogy to Howell's experiment is that Alice and Bob
must implement a time-bin system in order to ensure that the measurements they are
making are on modes produced from the same event.  Additionally, they must establish
beforehand, via an open channel, that if one of them measures the mode with the two
additional photons, it will correspond to a certain bit.  For example, if Alice measures
$n+2$ photons (implying Bob measures $n$), the bit will be a zero.  Correspondingly, if
Alice measures $n$ photons and Bob measures $n+2$, the bit will be a one.  The
measurement outcomes of which mode contains $n$ or $n+2$ photons are completely random
run-to-run.  Repeating this process will result in a perfectly correlated string of bits
between Alice and Bob, thus establishing a key for use as a one-time pad.  Noise-free
photon number-resolving detectors with up to $88\%$ efficiency have been experimentally
demonstrated at NIST \cite{nist}.  However, imperfect photodectors have been shown to
provide useful reconstruction of photon-number distributions as well \cite{dowling04}.

The security of the system is established in a manner completely analogous to Howell's
experiment; namely, Alice and Bob's measurement devices must consist of a Franson
interferometer \cite{franson}.  This detection scheme requires that Alice and Bob each
use an unbalanced Michelson interferometer, resulting in interference fringes due to the
path mismatch between the two modes they are measuring. It has been shown that the
Franson fringe visibility corresponds to a Bell-type inequality, which allows for
detection of an eavesdropper if there is a reduction in the fringe visibility
\cite{howellqkd,franson}.

Another straightforward application of the output state is to quantum metrology and
quantum lithography.  If we obtain a detection of exactly one photon at each detector
$D_{a}$ and $D_{d}$, thus telling us that $n=m=1$, we know with certainty the entangled
inner modes are in the state $\ket{\rm{inner}}=(\ket{3,1}+\ket{1,3})/\sqrt{2}$.  If this
state (in modes B and C) is then incident on a beam splitter, using the transformations
\cite{"barnettbook"},
\begin{eqnarray}
\hat{b}^{\dag}\rightarrow
\frac{\hat{b}^{\dag}+e^{i\theta}\hat{c}^{\dag}}{\sqrt{2}}~~,~~~\hat{c}^{\dag}\rightarrow\frac{\hat{b}^{\dag}-e^{i\theta}\hat{c}^{\dag}}{\sqrt{2}},
\end{eqnarray}
and for $\theta=\pi$ we obtain the $N=4$ $N00N$ state, $(\ket{4,0}+\ket{0,4})/\sqrt{2}$.
As discussed earlier, if this state is used to measure a path-length difference in a
Mach-Zehnder interferometer, it achieves a doubling in sensitivity compared to the
standard shot-noise limit.  Regarding use as a source for quantum lithography, proposed
by Boto et al, 4-photon $N00N$ states are predicted to achieve interference patterns of
the form $1+\rm{cos}(4\phi)$, where the phase $\phi$ corresponds to translation alone
the substrate \cite{boto}. This corresponds to a four-fold improvement in resolution
compared to the classical case, for which the pattern is of the form
$1+\rm{cos}(\phi)$.

\begin{figure}[t]
\includegraphics[width=6.5cm]{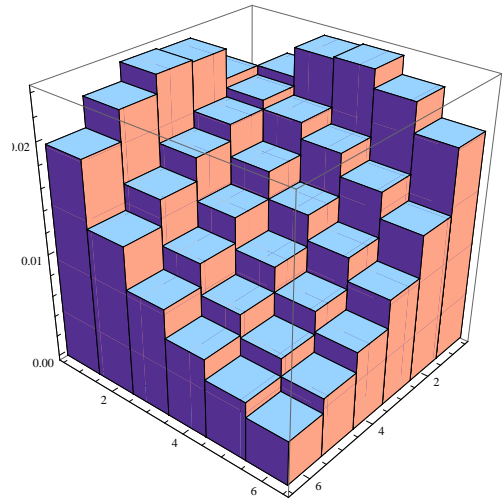}
\caption{\label{fig:epsart4} Probabilities of obtaining $n$ and $m$ photon states for
fixed $r=1.08$.  The most likely joint photodetection at detectors $D_{a}$ and $D_{d}$
is when each mode contains only one photon.  The vacuum $n=m=0$ term is the top diagonal
term, and is not visible because the $n=m=1$ term is more probable.  A joint
photodetection of $n=m=1$ leads to the entangled state $\ket{3,1}+\ket{1,3}$ between
modes $b$ and $c$, which when incident on a beam splitter leads to the $N=4$ $N00N$
state $\ket{4,0}+\ket{0,4}$.}
\end{figure}

Optimizing the gain $r$ such that we obtain the highest probability of obtaining a
measured output state of $n=m=1$ gives a quantatative prediction for how often the
desired state for quantum lithography will be heralded.  The optimal value is $r=0.66$.
However, we are also able to find values of gain such that the $n=m=1$ output state is
more likely to occur than the vacuum or any other $n=m$ output.  This is due to the
output state dependence on the number of photons in the modes, which is different from
standard two-mode squeezed vacuum, as previously mentioned.  Fig.~\ref{fig:epsart4}
shows the probabilities of obtaining a measurement of $n$ and $m$ at the two detectors.
The diagonal values are where $n=m$.  The inability to see the $n=m=0$ term is due to
the entanglement-seeding of the two OPAs.  The value of gain in this plot is $r=1.08$.
This value is easily obtainable \cite{grangier,sciarrino}.  Comparing the probabilities
of obtaining the $N=4$ $N00N$ state to that of a typical linear optics based scheme
\cite{leedowling}, we find that the dual OPA scheme produces the desired state more
frequently.  In Reference \cite{leedowling} the $N=4$ $N00N$ state is probabilistically
produced $3/64$ of the time.  Our state produces the same state at approximately $5$
times that rate.  This is due to the fact that the linear optical scheme relies on an
input state of $\ket{3,3}$, whereas our scheme requires that each crystal produces the
state $\ket{1,1}$, which is much more likely for OPAs.  Also, our scheme is able to
minimize vacuum contributions and shift the maximum probability to higher photon number,
as mentioned before, and as seen in Fig.~\ref{fig:epsart3} and Fig.~\ref{fig:epsart4}.

\section{Conclusion}

In conclusion, we propose a scheme that involves seeding two optical parametric
amplifiers with an entangled state input.  The amplification of this entangled input
state results in a four-mode entangled output state, which is useful in a variety of
applications.  Two of the modes may be measured, thus providing insight into what
entangled state the other two modes are in.  The heralded output state is a perfect fit
for quantum cryptographic purposes; analogous to polarization entangled quantum key
distribution as envisioned by Ekert. Additionally, due to parametric amplification, the
output state probabilities depend on the number of photons in the modes.  For
experimentally realistic values of gain, this allows for a high probability of obtaining
specific outputs, assuming photon number resolving detectors \cite{nist,dowling04}.
Specifically, the scheme produces heralded $N=4$ $N00N$ states when a triggered output
is incident on a 50:50 beam splitter.  This state can then be used for quantum
interferometry and lithography.  The setup employs three nonlinear crystals, all of
which can be identical, save that the seeding crystal needs to be pumped in the low-gain
regime.  The other optical tools needed are beam splitters and photodetectors.

\begin{acknowledgments}
We would like to acknowledge support from the Army Research Office, the Intelligence
Advanced Research Projects Activity, and the Defense Advanced Research Projects Agency.

\end{acknowledgments}

\bibliography{opapaper}

\end{document}